\documentstyle[12pt]{article}

\begin{document}


\hsize=6.15in
\vsize=8.2in
\hoffset=-0.42in
\voffset=-0.3435in

\normalbaselineskip=24pt\normalbaselines

\begin{center}
{\large \bf Synaptic polarity of the interneuron circuit controlling 
{\it C. elegans} locomotion}
\end{center}

\vspace{0.15cm}

\begin{center}
{Franciszek Rakowski$^{a}$, Jagan Srinivasan$^{b}$, Paul W. Sternberg$^{b}$, 
and  \\
Jan Karbowski$^{c,d,*}$ }
\end{center}

\vspace{0.05cm}

\begin{center}
{\it $^{a}$ Interdisciplinary Center for Mathematical and Computational 
Modeling, \\ 
University of Warsaw, 02-106 Warsaw, Poland \\
$^{b}$ California Institute of Technology, Division of Biology, Pasadena, 
CA 91125, USA \\
$^{c}$ Institute of Applied Mathematics and Mechanics, \\
University of Warsaw, 02-097 Warsaw, Poland \\
$^{d}$ Institute of Biocybernetics and Biomedical Engineering, \\
Polish Academy of Sciences, 02-109 Warsaw, Poland }

\end{center}


\vspace{0.05cm}

\begin{abstract}
{\it C. elegans} is the only animal for which a detailed neural connectivity 
diagram has been constructed. However, synaptic polarities in this diagram, and 
thus, circuit functions are largely unknown. Here, we deciphered the likely 
polarities of 7 pre-motor neurons implicated in the control of worm's locomotion, 
using a combination of experimental and computational tools. We performed single 
and multiple laser ablations in the locomotor interneuron circuit and recorded 
times the worms spent in forward and backward locomotion. We constructed a 
theoretical model of the locomotor circuit and searched its all possible synaptic 
polarity combinations and sensory input patterns in order to find the best match 
to the timing data. The optimal solution is when either all or most of the 
interneurons are inhibitory and forward interneurons receive the strongest input, 
which suggests that inhibition governs the dynamics of the locomotor interneuron 
circuit. From the five pre-motor interneurons, only AVB and AVD are equally likely 
to be excitatory, i.e. they have probably similar number of inhibitory and excitatory 
connections to distant targets. The method used here has a general character 
and thus can be also applied to other neural systems consisting of small 
functional networks.

\end{abstract}




\newpage


\vspace{0.1cm}

\noindent ${*}$ Corresponding author at: jkarbowski@duch.mimuw.edu.pl

\vspace{1.1cm}

\newpage

\noindent {\Large \bf Introduction}

{\it Caenorhabditis  elegans} nematode worms possess a very small nervous system 
composed of only 302 neurons connected by about 5000 chemical synapses and 3000 gap 
junctions \cite{white}. Because of its smallness a precise map of neural connections 
was possible to construct \cite{white,chen}. This places {\it C. elegans} in a unique 
position among all other animals \cite{varshney}, for which we have at best only 
rudimentary connectivity data to test various concepts regarding neural wiring 
and function \cite{chen,cherniak,kaiser,karb2001,karb2003,chklovskii}. However, 
despite this achievement we still have a very limited knowledge about the nature 
of most of the worm's synaptic connections, i.e. whether they are excitatory or 
inhibitory.

The simplicity of the {\it C. elegans} nervous system does not preclude these 
worms from executing various non-trivial behaviors such as locomotion, feeding, 
mating, chemotaxis, etc \cite{hobert,debono}. To understand the neural basis of these 
behaviors requires some information not only about the pattern and strength of the 
connections but also about the type of their synapses. The same neural circuit can 
perform different functions depending on the signs of synaptic polarities it contains. 
Specifically, circuits in which excitatory synapses dominate can sometimes become 
epileptic. On the other hand, networks with only inhibitory connections could be silent, 
and therefore in many situations useless. Thus, it may seem that some sort of an 
intermediate regime is necessary for a proper functioning of the nervous system 
\cite{vreeswijk}. For example, it was proposed that mammalian cortical networks operate 
in a dynamic state in which excitation is effectively balanced by inhibition 
\cite{haider,vogels}, although anatomical number of excitatory connections dominates 
over inhibitory in the cerebral cortex \cite{defelipe}. For nematode worms, a
similar issue has been addressed only sporadically. On a modeling level, in the
context of a tap withdrawal circuit \cite{wicks}, and experimentally, studying
genes that influence the ratio of excitatory to inhibitory signaling \cite{jospin}.
We think that this topic deserves more attention both theoretical and
experimental, if we are to understand the functioning of worm's circuits 
\cite{sengupta,gray,ha}.

Movement direction in {\it C. elegans} is governed by 5 distinct locomotory 
command interneurons (AVB, PVC, AVA, AVD, AVE), each in two copies (left and right).
All of these 10 interneurons directly connect a downstream group of dorsal and 
ventral body wall excitatory motor neurons \cite{chalfie}. The topology of connections
between the command interneurons is well known \cite{white,chen}, however, their 
synaptic polarities are not. Conventional thinking is that AVB and PVC control 
forward motion, while AVA, AVD, AVE control backward motion \cite{chalfie}. This 
reasoning is based on the fact that the former interneurons connect mainly motor 
neurons of type B (experimentally shown to be critical for forward locomotion 
\cite{haspel}), whereas the latter interneurons connect mostly type A motor neurons 
(ruling the backward motion \cite{haspel}). However, this simple locomotory picture, 
relying on a single neuron function doctrine may turn out to be too simplistic. Indeed, 
many laser ablation experiments show that removal of both AVB and PVC reduces forward 
motion, but does not abolish it completely (see below). Similarly, worms lacking 
the ``backward'' interneurons AVA, AVD, and AVE exhibit a comparable frequency of 
reversals as intact wild type worms \cite{piggott}. Moreover, the major backward 
interneuron AVA makes also connections (both synaptic and gap junctions) with 
the forward B motor neurons \cite{chen}. Thus, perhaps the decision to move in 
a particular direction is generated by a collective activity of all command 
interneurons, rather than by an activity of a particular interneuron or a 
particular connection.

Our aim is to investigate the problem of decision making for movement direction
in {\it C. elegans} on the level of its interneuron network. The main question we 
pose is how two antagonistic behaviors, i.e. forward and backward motions, can
be controlled by the same circuit of mutually coupled pre-motor interneurons.
A strictly related to this question is the problem of synaptic polarities of these 
interneurons and the input pattern they receive. Specifically, by applying structural 
perturbations to the circuit we want to determine, using mathematical modeling, 
which combination of synaptic polarities (together with an input pattern) gives 
the best match to the experimentally observed locomotory output of {\it C. elegans}. 
This knowledge allows us to answer a question about a relative influence of inhibition 
and excitation in the command interneuron circuit. Moreover, this approach provides an 
insight about a degree of interneuron collectiveness in choosing the direction of motion.

\newpage

\noindent {\Large \bf Results}

\vspace{0.5cm}

\noindent {\large \bf The command interneuron circuit for {\it C. elegans} 
locomotion}

To simplify data analysis and mathematical modeling we grouped the left and right 
members of each locomotor circuit neuron as one model neuron. Thus, in our 
circuit controlling worm's motion there are 5 command interneurons, one distinguished 
upstream polymodal sensory interneuron called ASH, and a modulatory neuron DVA 
(Fig. 1A). A recent study \cite{li} indicates that DVA plays a role of a sensory
neuron in locomotion. We included this neuron explicitly in the circuit, since it has 
direct connections with body wall motor neurons \cite{chen}, similar to 5 command 
interneurons. Because of this similarity, we want to investigate whether DVA can 
serve additionally as a command interneuron.

The neurons in the circuit are modeled as a single passive compartment with leak
conductance. Connections between neurons are either by chemical synapses (of unknown 
polarity) or by electric synapses known as gap junctions. Chemical synapses transmit 
signals using graded transmission. The strength of the connection between two 
arbitrary neurons is proportional to the number of anatomical contacts between them 
determined from the empirical data \cite{chen}. Additionally, each pre-motor interneuron 
receives a constant in time excitatory input from upstream (mostly head) neurons, 
which can be either weak or strong (this is variable in the model). Overall, our
model captures long-term averages in neural activities that are associated with
average locomotory output in {\it C. elegans}. All the assumptions made in the model 
and equations describing activities of all neurons are presented in Materials and 
Methods and in Supplementary Information. The main parameters used in the model are 
described in Table 1.

In an intact circuit for wild type (WT) worms fluctuations in interneuron activities
control forward and backward motion, and the distribution of these activities 
determines the relative proportion of forward and backward motion (roughly 3 to 1). 
We wanted to perturb the system and investigate its corresponding output by performing 
laser ablations of selected neurons in the locomotory circuit. We reasoned that 
a gradual removal of neurons from this circuit (Fig. 1B; see Materials and Methods) 
would not only affect its physical structure, but also would redistribute the 
remaining neurons activities, which in turn, should modify the worm's locomotory 
behavior. In particular, the pattern of interneuron activities in the circuit 
should change, altering the ratio of times spent in forward and backward motion. 
Thus, associating experimental average times of forward and backward locomotion 
for every ablation type with changes in the average activity levels of the circuit
model for fixed sensory inputs can allow us to predict synaptic polarities of 
the interneurons.

\vspace{0.5cm}

\noindent {\large \bf Experimental results}

We performed single, double, triple, and quadruple ablations in the interneuron 
circuit. In total, we generated 17 types of ablation and recorded corresponding mean 
times the worms spent in forward ($T_{f}$) and backward ($T_{b}$) motion, as well as 
in stopped phase ($T_{s}$). These experimental results are presented in Table 2. 
From all the ablations executed, only removal of ASH and PVC neurons increase the 
time spent in forward locomotion in relation to wild type. This is an indication that 
these particular interneurons have a definitely negative influence on the forward 
direction, and for that reason are likely to be inhibitory (see below). All other 
removals have a detrimental effect on the duration of forward motion, even those 
traditionally associated with backward motion (AVA, AVD). In particular, the ablation 
of AVA has the most dramatic effect on $T_{f}$, leading to its 10 fold reduction in 
comparison to wild type, although the reversal frequency increases only mildly by 
a factor of 2. Moreover, the AVA ablated worms, including their multiple ablations, 
spent a lot of time not moving (stopped phase), far more than WT and worms with other 
types of ablation. For instance, for the combined ablation ASH+AVA+AVB, we find the 
largest stopped mean time of 1.16 sec.

Worms with multiple ablations reverse roughly as frequently as worms with single 
ablations (Table 2). Generally, the frequency of direction reversals does not correlate 
well with the average time worms spent in forward motion (Table 2). For example, 
worms with killed ASH reverse approximately as often as worms with removed AVD, 
despite the fact that ASH worms spent 3 times more time in forward motion. 
Similarly, worms with almost equal $T_{f}$ (about 0.9 sec), i.e. AVB+PVC and 
AVA+AVB+PVC, differ in reversals by a factor of 2.5.

An interesting result is that ablating the modulatory neuron DVA causes
a sharp decline in the forward motion timing in comparison to WT, and a two-fold 
increase in reversals (Table 2). Also, the combined ablations of DVA with 
PVC and AVB show a similar property. This clearly suggests that this neuron has 
a significant influence on the interneuron circuit output, which could be more 
than just its sensory modulation.

\vspace{0.5cm}

\noindent {\large \bf Theoretical results}

In our circuit model there are 7 neurons (5 pre-motor interneurons, DVA, and ASH), 
each of them can be either excitatory or inhibitory. Thus, there are $2^{7}= 128$ 
possible copies of the circuit associated with synaptic polarities, i.e. the sign 
of $\epsilon_{i}$ (for $i=1,...,7$). Two examples of the polarity copies are:
(i) $\epsilon_{ASH}= -1$, $\epsilon_{AVB}= -1$, $\epsilon_{PVC}= -1$, 
$\epsilon_{DVA}= -1$, $\epsilon_{AVA}= -1$, $\epsilon_{AVD}= -1$, 
$\epsilon_{AVE}= -1$, and (ii) $\epsilon_{ASH}= 1$, $\epsilon_{AVB}= 1$, 
$\epsilon_{PVC}= 1$, $\epsilon_{DVA}= 1$, $\epsilon_{AVA}= 1$, $\epsilon_{AVD}= 1$, 
$\epsilon_{AVE}= 1$, which correspond to all inhibitory and all excitatory neurons, 
respectively. Additionally, each of the 6 pre-motor neurons (excluding ASH) receives 
an upstream sensory input coming mostly form the head, except for PVC for which it 
comes predominantly from the tail. The sensory input is represented by a binary 
variable $z_{i}$ that can have two values: 0 for a weak input, and 1 for a strong input
(see Eq. 7 in Materials and Methods). 
Consequently, every polarity copy can be found in additional $2^{6}= 64$ activity 
or input configurations. This implies that, in total, our circuit model has 
$2^{7}\cdot 2^{6}= 8192$ distinct polarity-input configurations.

For each possible configuration (i.e. synaptic polarity and an upstream input) of 
the circuit we performed 17 ``computer ablations'' analogous to the experimental 
ablations shown in Table 2, by setting $\epsilon_{i}= 0$ if the neuron with an index 
$i$ was removed. This procedure removes all the connections (synaptic and electric) 
coming out of this ablated neuron. For each ablation we computed the fraction of 
time corresponding to forward motion, i.e. $T_{f}/(T_{f}+T_{b})$, using Eq. (8). 
Thus, we generated 18 theoretical fractional times associated with every circuit 
configuration (17 types of ablation plus WT) and computed their Euclidean distance 
(ED) to the experimental values in Table 2 (see Materials and Methods for a more 
detailed description). The configuration with the smallest ED value corresponds to 
the optimal solution that predicts synaptic polarities and the pattern of the upstream
input. Our strategy was to vary the level of this input ($\sigma$, $\kappa$), and 
for each level to find optimal values of synaptic and gap junction conductances 
($q_{s}$, $q_{e}$), together with the system noise amplitude ($\eta$), which give 
the best fit of the theoretical $T_{f}/(T_{f}+T_{b})$ to the experimental data.

\vspace{0.3cm}
\noindent {\bf Winning configurations.} 

We find that the best match to the experimental data is achieved in the case when 
all 7 neurons in the circuit are inhibitory (Tables 3-5). The winning synaptic 
polarity configuration is associated with the combination number 1, which gives 
the best (lowest) ED= 0.363, and the largest correlation with the data points, 
which is 0.743 (Table 3). The distribution of winning values of the upstream input 
$z_{i}$ is non-homogeneous, and nonzero only for AVB and PVC neurons, implying that 
much sensory excitation comes to the neurons controlling directly forward motion 
(Table 3). In Fig. 2 we display a comparison of theoretical and experimental values 
of $T_{f}/(T_{f}+T_{b})$ across different ablations for this winning combination of 
synaptic polarities. In general, there is a good fit of the theoretical points to 
the data (correlation about 0.74).

The second place among all synaptic polarities is taken by the combination 
$\#$ 17, for which all neurons, except AVB, are inhibitory. This combination 
has ED value very close to that obtained by the winning combination $\#$ 1 
(Table 3). Moreover, these two combinations appear the most often among the winners, 
also for other choices of parameters describing the sensory input (Tables 3-6). 
The third place is taken by a combination with the number 11, in which only two 
neurons are excitatory: AVD and DVA. This combination also appears quite often 
among the leading polarities, but its ED is in some separation from the two winning 
synaptic combinations 1 and 17.

\vspace{0.3cm}
\noindent {\bf Optimal solution depends both on synaptic and input configurations.} 

The degree of the match between theory and experiment, i.e. ED, depends not only
on the pattern of synaptic polarities but also on the pattern of incoming sensory 
input ($z_{i}$) to the network (Fig. 3). We noticed that we obtained better fits 
if we allowed the interneurons to receive a heterogeneous input from upstream neurons.
A given synaptic polarity configuration can produce a slightly different locomotory 
output (slightly variable ED) depending on how many, and which, interneurons receive 
a strong input (Fig. 3). However, despite this subtlety the overall emerging picture 
is such that the smallest ED values are associated with configurations dominated by
inhibitory connections, while the largests ED correspond to networks dominated by 
excitatory synapses, regardless of the input pattern. Among the configurations
with the lowest ED, the most optimal are those with a moderate number (typically 2 
or 3) of interneurons receiving strong input. The highest value of ED is about 2.0 
and it occurs for synaptic configurations with prevalent excitatory connections,
with only a minor influence of the input pattern (Fig. 3). These results suggest that
the Euclidean distance ED is much more affected by the pattern of synaptic polarities 
than by the input pattern.

\vspace{0.3cm}
\noindent {\bf Likelihood estimates of interneuron synaptic polarities.} 

To quantify the likelihood of a given synaptic polarity among leading
combinations we associate with each neuron a probability that it is inhibitory, 
for each input value $\sigma$ and ASH activity level (see last columns in Tables 3-6).
This probability is defined here as a fraction of times the $\epsilon= -1$
appears in the row for each neuron. One can notice that some of the locomotory 
interneurons, such as AVE, AVA and PVC, are inhibitory with probabilities that 
are close to 1, regardless of the input magnitude. The polymodal neuron ASH is 
also in this category. For the rest of the pre-motor interneurons these probabilities 
are not that high, but nevertheless are at least $\ge 0.5$. We also computed an 
average probability that a given neuron is inhibitory across different input levels 
coming to the locomotory circuit (average values for all Tables 3-6). These average 
probabilities are: 1 for ASH, 0.875 for AVA, 0.5 for AVB, 0.5 for AVD, 0.938 for AVE, 
0.656 for DVA, and 0.719 for PVC. Thus, from the whole group of 7 neurons investigated 
here and implicated in locomotion control, only AVB and AVD have about equal chances 
to be excitatory and inhibitory. 

One might wonder if these probability estimates hold if we include more winning 
combinations, not just eight as in Tables 3-6. Including 20 leading combinations 
for the winning input values $\sigma=8$ mV and $\kappa=0.6$ (corresponding to Table 3), 
gives qualitatively a similar picture. Again, the neurons AVB and AVD have the 
highest likelihood of being excitatory, although inhibitory polarities of these 
interneurons have the lowest ED values (Fig. 4). Taken together, the above results 
strongly suggest that the majority of interneuron connections are inhibitory.

\vspace{0.3cm}
\noindent {\bf Dependence of ED on model free parameters.} 

The Euclidean distance ED between theoretical and experimental fractions of time
spent in forward locomotion depends in a non-monotonic manner on model free 
parameters: $q_{s}, q_{e}, \sigma, \kappa$, and $\eta$. The first four parameters
are neurophysiological in nature and determine the overall balance between currents 
flowing in the interneuron network. The last parameter, $\eta$ characterizes a shape
of the transfer function between neural activities and behavioral locomotory output,
or in other words it characterizes a network noise level.

In Fig. 5 we show a dependence of ED on synaptic and electric conductances ($q_{s}$ 
and $q_{e}$) in the system. In general, ED has a minimum for a narrow range of these 
conductances, indicating that an optimal solution exists for this range. A qualitatively 
similar picture emerges also for the rest of the free parameters. Typically, the parameters 
controlling the strength of the upstream input, i.e. $\sigma$ and $\kappa$, should be in 
some intermediate range to reach a minimal value of ED. For instance, the winning polarity 
combinations for $\sigma= 4$ and $\sigma= 12$ mV have larger ED  than that for 
$\sigma= 6$ or 8 mV (Tables 3-6). The same is true for the value of $\kappa$, 
characterizing ASH activity (see Materials and Methods), with the optimal $\kappa= 0.6$ 
for each $\sigma$ level.

In Fig. 6 we show ED as a function of noise level $\eta$. Again, either too strong or 
too weak $\eta$ increases ED, and there is an optimal value of this parameter for which 
ED has a minimum.

\newpage

\noindent {\Large \bf Discussion}

\vspace{0.4cm}
\noindent {\bf The main findings.} 

Using a combination of experimental (laser ablations) and computational
(circuit model and optimization) tools we were able to decipher the likely 
synaptic signs of the interneurons composing the small network commanding 
{\it C. elegans} locomotion \cite{white,chalfie}. It turns out that probably 
most of these neurons, i.e. synapses they sent, are inhibitory. In particular, 
the average probabilities that a particular interneuron is inhibitory are: 
0.875 for AVA, 0.5 for AVB, 0.5 for AVD, 0.938 for AVE, 0.719 for PVC, and 
0.656 for DVA. These numbers suggest that although some of the connections 
coming out of these neurons might be excitatory, the majority of the connections 
are clearly inhibitory.

Because of a suppressing nature, the command interneuron circuit must receive 
a sufficient amount of excitation from upstream (in large part sensory) neurons 
to be functional, i.e. to appropriately activate downstream motor neurons. 
Our computational results indicate that the behavioral data are best explained 
if the command circuit receives a mixed, heterogeneous input (denoted by $X_{i}$
in our model; Eqs. (6) and (7)). The best fit to the data is obtained if the largest 
excitation comes to forward interneurons AVB and PVC (Table 3; Fig. 3). In this 
sense, the existence of sensory stimulation is an important factor for directional 
motion generation, which is in general agreement with the experimental findings 
\cite{piggott,zheng}.

\vspace{0.4cm}
\noindent {\bf Role of interneuron gap junctions in locomotion.} 

How are downstream motor neurons activated given an inhibitory nature of
synapses in the pre-motor interneuron circuit? This probably occurs due to
strong gap junction couplings between two major interneurons, AVB and AVA,
and the downstream motor neurons of type A and B. Our results suggest that
the worm moves forward because AVB receives a stronger upstream input than
the ``backward'' interneurons. AVB then excites downstream B motor neurons 
via strong gap junctions. The chemical synapses from AVB to B are weak and
thus non-significant (Table 7). Therefore the sign of these synapses is 
irrelevant for forward motion, and AVB does not necessarily have to be 
excitatory.

The issue with backward motion is more subtle. Recent calcium imaging studies 
show that AVA is active during backward movement \cite{faumont,chronis}. 
How can one explain this? The likely answer lies again in the strong gap junction 
connections between AVA and A motor neurons, and between AVB and B motor neurons 
(Table 7). Specifically, during backward motion AVA, due to its large sensory 
input, synaptically inhibits other interneurons including AVB, but at the same 
time excite downstream A motor neurons via strong electric coupling. Thus AVB 
sends less excitation to B motor neurons via its strong gap junctions than does 
AVA to A neurons. Consequently, the activity of A prevails over the activity of 
B neurons (i.e. $E_{b} > E_{f}$ in our model), and the worm moves backward, 
even with inhibitory synapses in the locomotor circuit.

\vspace{0.4cm}
\noindent {\bf Relative importance of the patterns of synaptic polarity and 
sensory input on the results.} 

Our results indicate that locomotory output of the interneuron circuit depends
on its synaptic polarities as well as on sensory input pattern. This can be seen
in Fig. 3, where the Euclidean distance ED characterizing the degree of the match 
between the theory and the data is displayed as a function of synaptic and input
configurations. From these two main factors affecting ED, the synaptic polarity
pattern seems to be a more important determinant. This is because the lowest values 
of ED are generally associated with circuit configurations in which synaptic inhibition 
dominates, whereas the highest ED's are related to mostly excitatory connections (Fig. 3). 
On the other hand, the sensory input pattern does not exhibit such a simple general 
trend. More precisely, increasing the amount of strong input in the network does not 
lead to an explicit decrease (or increase) of ED, but rather to its fluctuations. 
Instead, ED assumes the smallest values for a moderate number of interneurons 
receiving strong input, which is usually 2 or 3 (Fig. 3).

\vspace{0.4cm}
\noindent {\bf Properties of ASH and AVB neurons strongly affect the results.}

From all neurons present in our locomotory circuit, two neurons ASH and AVB
are particularly important for the network performance. This is strongly related
to the question of specificity in synaptic polarity and input pattern. The map in 
Fig. 3 shows that the synaptic polarity of ASH is strongly correlated with the level 
of ED. In most cases, synaptic configurations in which ASH is inhibitory decrease ED, 
and increase it if ASH is excitatory. The exceptions are configurations with small 
inputs, where the opposite takes place. Thus, for the most part the synaptic polarity
of ASH should be inhibitory in terms of the network optimality.

AVB neuron, which provides the main signal to the downstream motor neurons controlling
forward locomotion, is also critical for the optimal solutions. The strength of the 
sensory input this neuron receives strongly correlates with ED (Fig. 3). Generally,
strong input in AVB decreases ED, whereas weak input in AVB increases ED, and this
effect is essentially independent on synaptic polarity configurations (Fig. 3). This 
suggest that on average AVB should receive a strong excitatory input coming from 
upstream neurons.

If we combine the action of both ASH and AVB we obtain an interesting picture. ED
values are the smallest for cases in which ASH is inhibitory and AVB receives a strong
input (Fig. 3). In an opposite case, when ASH is excitatory and AVB gets a weak
input, ED values are the largests.

\vspace{0.4cm}

\noindent {\bf Dependence of network performance on model free parameters.}

Performance (ED) of the interneuron circuit depends also on several
neurophysiological parameters and on the level of noise in the system.
Generally, these parameters require fine-tuning to obtain the the best fit
between the theory and the data. Synaptic conductance $q_{s}$ and gap
junction conductance $q_{e}$ determine the relative strength of synaptic
and electric connections. Optimal values of these parameters that give
the best model performance are always in a physiological range, and yield
very similar values $\sim 0.1$ nS. However, ED is more sensitive to changes
in synaptic conductance $q_{s}$ than to changes in $q_{e}$ (ED has a broad
minimum as a function of $q_{e}$; Fig. 5), which suggests that precise
values of synaptic conductance are more important.

ED has always a minimum as a function of noise amplitude, regardless of the
values of other parameters (Fig. 6). This optimal value of noise is in the range
0.7-1.1 mV (Tables 3-6), which seems to agree with the magnitude for voltage 
fluctuations obtained from electrophysiology \cite{piggott}. The optimality of
the system performance for some finite level of noise resembles a phenomenon
known as a stochastic resonance, which is a ubiquitous mechanism in many physical
and biological systems \cite{mcdonnell}.

The circuit performance not only depends on the pattern of the sensory input 
(see above), but also on the strength of this input (Tables 3-6). There exists
some optimal level of upstream excitation for which ED has a minimum. Either
increase or decrease of this level has a negative influence on the system 
performance. This non-monotonic dependence can be explained in the following
way. When the incoming sensory input is too strong its contribution to
an interneuron voltage is much bigger than the contributions coming from
synaptic and gap junction couplings. Thus manipulation of parameters associated
with connectivity does not change the network output, and one cannot improve
the system performance. On the other hand some minimal input is required
to stimulate the network, so the signal flows to the motor neurons.

The level of excitation in ASH neuron, controlled by $\kappa$, can be also
viewed as a measure of an additional input coming to the interneuron circuit.
Therefore, a non-monotonic dependence of ED on $\kappa$ should not be a surprise,
which is what we observe in our computational analysis.

\vspace{0.4cm}

\noindent {\bf Asymmetry in forward and backward interneuron activities
determine the likely direction of motion.} 

Our results indicate that a decision to move in a particular direction can be made 
on a small circuit level composed of the 6 pre-motor interneurons (including DVA). 
Specifically, the output from these interneurons is fed to the two types of
body wall motor neurons (B and A) controlling forward and backward motions, 
whose relative asymmetric activities ($E_{f}$ and $E_{b}$ in our model) determine 
the likely direction of worm's movement. This simple decision making mechanism 
can explain 74 $\%$ of the correlations between the experimental data and 
computational results (see Table 3). Moreover, this behavioral picture is 
consistent with the findings in a recent experimental study \cite{kawano}, in 
which it is shown that the imbalance between activities of A and B motor neurons 
is a likely scenario for the selection of worm's motion direction. However, in
contrast to these authors the imbalance between A and B neurons in our model
is caused not so much by a strong AVA-A gap junction coupling, but by the
asymmetric upstream excitatory input to command interneurons (in the winning
combinations, AVB and PVC neurons receive much stronger upstream excitation 
than the rest of the circuit).

\vspace{0.4cm}

\noindent {\bf Qualitative interpretation of ablation data.}

The interesting experimental result concerning single-neuron ablations is that 
removal of certain interneurons causes an increase in forward motion timing, 
while removal of others leads to its dramatic decrease. Specifically, only 
killings of ASH and PVC neurons increase significantly the time $T_{f}$ 
spent in forward locomotion. In relation to ASH, this suggests an important role 
of the sensory input. There are two surprises here. First, PVC was thought as 
promoting forward locomotion \cite{chalfie}. Second, given that the sensory ASH 
neuron makes only weak or intermediate synaptic connections with the command 
interneurons (with all ``backward'' interneurons and AVB; see Table 7), it should 
not have such a strong influence on the motion characteristics. The solution of 
these puzzles is that PVC and ASH are probably inhibitory, i.e. the strongest 
synapses connecting these neurons with other postsynaptic targets should be 
inhibitory. Additionally, ASH should be highly depolarized in order to significantly 
downregulate the locomotory (mostly backward) interneurons via its synapses.
Recent calcium imaging indicates that ASH to AVA synapse is likely excitatory
\cite{chronis}. This experimental result does not necessarily contradict our
result regarding the negative ASH polarity. Indeed, our results concern only 
an average total polarity of a given neuron, and it is possible that some of
its weak synapses can have a reverse polarity in relation to the strongest. 
In fact, the ASH to AVA synapse is relatively weak (third in strength out of four 
coming out of ASH; see Table 7), and it could be dominated by stronger inhibitory 
synapses to other neurons.

Removal of AVA interneuron causes a large reduction in $T_{f}$, despite the fact 
that this neuron belongs to the ``backward'' locomotory circuit, and one might 
naively expect that it effectively prohibits forward motion. Moreover, single and 
multiple ablations associated with AVA cause an increase in stopped time (Table 2). 
This suggests that removal of AVA decreases a difference between activities of 
forward and backward motor neurons, i.e. $E_{f}-E_{b}$ may become comparable with 
a threshold for movement initiation $\Delta$ (see Supplementary Information). This 
in turn may suggest that when AVA is absent, backward motor neurons are more active. 
Taken together, these results imply that the overall synaptic effect of AVA is most 
likely inhibitory.

The ablation results for the DVA neuron indicate that it plays a more 
significant role in the locomotory circuit than just its passive modulation. 
From Table 2 it is evident that killing DVA has one of the biggest impacts on 
$T_{f}$. Based on this, we hypothesize that DVA might serve also as a command 
neuron in generation of forward locomotion, which is a novel function for
this neuron.

The case with multiple ablations is more complicated. These type of ablations 
do not have an additive property, i.e. removal of more neurons does not necessarily 
lead to a progressive drop in the forward motion timings. For example, double AVB 
and PVC ablation has $T_{f}= 0.91$ s, but additional removal of AVD actually 
increases $T_{f}$ to 1.33 s (Table 2). The latter may seem paradoxical, however 
one has to remember that backward neurons do not act in isolation, but participate 
in the whole interneuron network activity, and thus indirectly also influence 
the forward motor neurons. Apart from that, interneurons interact among themselves
both synaptically (nonlinear in nature) and via gap junctions (bidirectional
in nature). This may additionally mask a single interneuron contribution to 
the locomotory output of the circuit. As a result it is very difficult to predict 
in advance the effect of any particular ablation on worm's locomotory characteristics 
in the case of multiple ablations. For this, one needs to perform detailed computations 
on a network level, as was executed in this study.

\vspace{0.4cm}

\noindent {\bf Collective, mutually inhibitory interactions between command
interneurons underlie {\it C. elegans} direction of locomotion.}

These ablation results suggest that a picture in which a single neuron or a single
connection control a specific behavior, advocated in several former studies 
\cite{gray,chalfie}, may be oversimplified. Instead, our findings support 
the idea that behavioral (locomotory) output depends to a large degree on a 
collective activity of neurons comprising the ``functional circuit'' \cite{zheng}. 
That is, the same neuron can participate to some extent in opposite behaviors. 
Obviously, some neurons or connections in the functional circuits may be more 
dominant than others for a particular behavior, but the presence or absence of 
a particular neuron in the circuit is generally not critical for its operation. 
This partial redundancy in neural function is probably evolutionary driven to 
ensure a robust circuit performance.

Similarly, none of the interneuron ablations, either single or multiple, 
abolishes the worm's movement or body oscillations completely. This clearly 
indicates that none of the interneurons alone is a Central Pattern Generator,
which again speaks in favor of collective rather than individual interneuron
dynamics as a determinant of locomotion.

Our main result that the pre-motor interneuron circuit has mainly inhibitory 
synapses is qualitatively similar to two earlier \cite{wicks,zheng} and 
one recent \cite{qi} study. Wicks et al \cite{wicks} investigated computationally a 
tap withdrawal circuit in {\it C. elegans} and predicted that most interneurons 
composing it were inhibitory \cite{wicks}. That study concluded that PVC and AVD 
interneurons were probably excitatory. Our results suggest that AVD is equally
likely to be inhibitory as excitatory, whereas PVC with high probability should be
inhibitory. The possible sources of the discrepancy can be that Wicks et al 
\cite{wicks} used an older incomplete connectivity data for the pre-motor 
interneurons \cite{white}, did not include AVE neuron, and used a little
different set of neurons in their analysis. In another study, Zheng et al 
\cite{zheng} hypothesized that the locomotory interneuron circuit should act 
as an inhibitory switch in order to explain qualitatively data on motion direction 
transitions. A recent experimental work \cite{qi} also suggests that the pre-motor 
interneurons should use inhibition as a main synaptic signaling.

An interesting question is which neurotransmitters mediate inhibitory interactions
between interneurons. The most likely neurotransmitter between interneurons is
glutamate. In mammalian brains, it is known to be exclusively an excitatory
signal, since AMPA and NMDA postsynaptic receptors conduct mostly Na$^{+}$
and K$^{+}$ with an effective reversal potential around 0 mV. However, in
{\it C. elegans} the situation is more complicated because locomotor interneurons
contain apart from these receptors, also GluCl postsynaptic receptors \cite{brockie}.
These channels are gated by Cl$^{-}$ (with large negative reversal potentials),
and therefore mediate inhibition to postsynaptic cells \cite{brockie}. Specifically, 
the currents associated with GluCl receptor have been observed in the AVA 
interneuron \cite{mellem}, and they may also exist in other interneurons.

Given these two types of postsynaptic receptors, it is possible that a single
interneuron can have both excitatory and inhibitory synapses on distinct 
postsynaptic targets. In this case, the synaptic polarities associated with each
neuron in our study have an average character. More precisely, the determined
probabilities that a given neuron is inhibitory are the fractions of inhibitory
connections that the neuron makes with other postsynaptic neurons. Thus, for
example, the inhibitory probability 0.5 found for AVB indicates that this neuron
sends out about 50$\%$ inhibitory and 50$\%$ excitatory synapses.

\vspace{0.4cm}

\noindent {\bf Our computational model and its extension.}

The theoretical approach in this paper blends a traditional neural network
modeling with a probabilistic method for relating network activity to
behavioral data. In particular, we envision the nematode locomotion
as a three state system, one state for forward, second for backward motion,
and third state for no motion. In this system there are transitions between 
the states that are caused by intrinsic relative activities of A and B motor neurons 
($E_{b}$ and $E_{f}$), as well as by the system noise ($\eta$ in Eq. 8). 
Note that many ablations in Table 1 have the ratio $T_{f}/T_{b}$ close to unity, 
which in terms of our model implies that
$(E_{f}-E_{b})/\eta \ll 1$, i.e. for these ablations a stochastic influence of 
the environment is bigger than the relative activities of A and B motor neurons. 
This is interesting and shows that sensory noise gains in importance as we remove 
more neurons from our circuit. This may also suggest that some of the interneurons 
act as filters for the environmental noise.

Our combined approach allows us to achieve a concrete goal, which is the prediction 
of synaptic polarities for the well defined locomotory interneuron circuit and the 
determination of the likely sensory input pattern. This prediction does not depend 
on a precise form of the transfer function between neural activation and locomotory 
output (compare Eqs. 15 and 16 in the Supplementary Information). In this sense 
our results are robust.

The knowledge of the probable synaptic polarities of the command 
interneurons may have a positive impact on future modeling studies of 
{\it C. elegans} locomotion. We hope, that this will enable more realistic 
simulations of the neuronal dynamics that can extend the scope of testable 
predictions of the current locomotory models \cite{karb2006,karb2008,bryden}. 
Our method of determining synaptic polarities, which combines structural 
perturbations with the computational modeling, is sufficiently general that 
can be also applied to other small functional neural systems in which synaptic 
polarities are unknown. However, it is important to keep in mind that our
model, as every model describing biology, is subject to several assumptions 
(see the list in the Methods), and clearly has some limitations. The model 
does not include several subtle neurophysiological features. For example, 
a possibility that an individual neuron might release multiple neurotransmitters 
of similar importance, or that neuromodulators might provide an extra synaptic 
input throughout the network. These features can be addressed in future more
detailed investigations.

\newpage

\noindent {\Large \bf Materials and methods}

The ethics statements does not apply to this study.

\vspace{0.5cm}

\noindent {\large \bf Collection of experimental data}

\noindent {\bf Strain maintenance.} For our automated locomotion analysis, 
we cultured {\it C. elegans} at 20 $^{o}$C on NGM plates seeded with the OP50 
using standard methods \cite{brenner}.

\noindent {\bf Automated worm tracking and data extraction.}
Worms tested by automated tracking were continuously cultured on E. coli OP50. 
For assaying various parameters of locomotion, 10 cm non-seeded NGM plates 
were used. These NGM plates used for recordings were equilibrated to 
20 $^{o}$C for 18-20 hours. After ablations of individual neurons, the worms 
were placed on plates with $E.coli$ as a food source to recover. Ablated worms 
and mock controls were tested within 72 hours of the L4 molt. They were then 
transferred to assay plates containing no food. After 5 minutes of acclimatization 
on these plates, the worms were video taped for 5 minutes. Data presented in this 
paper represent the locomotory behavior of worms when they were exhibiting 
``area restricted search behavior''. Data extraction and processing was done 
using image processing and analysis software as previously described \cite{cronin}. 
From each video recording of 5 minutes, we used the middle 4 minutes, and used the 
software to derive values for times in forward and backward locomotion, as well as 
reversal frequencies. In our software, we used a velocity threshold for motion 
detection. Specifically, if the nematode velocity was below 0.05 mm/sec, we classified 
this as stopped time or ``no motion''. Every change of velocity direction that was above 
this value was classified as a reversal. The average time spent in the stopped phase
is $T_{s}$. The time $T_{f}$ the worm spent in forward motion is defined as an 
average duration of time counted from a moment of moving forward to stopping.
Similarly, the time $T_{b}$ spent in backward motion is an average of times
from the initiation of backward movement to stopping. Generally, because of
many reversals, the sum of the times $T_{f}, T_{b}$, and $T_{s}$ is much smaller
than the recording time of 4 minutes. The numerical values of $T_{f}, T_{b}, T_{s}$
provided in this paper are population averages. All incubations and recordings 
were done in a constant temperature room at 20 $^{o}$C.

\noindent {\bf Laser ablations.} For all species tested, we used the L1 larva 
stage for our ablations.

\vspace{0.5cm}

\noindent {\large \bf Description of the command interneuron circuit model}

\vspace{0.2cm}

\noindent {\bf List of the assumptions used to construct the model.} 
We make the following major assumptions in the theoretical model:

\noindent 1) In the interneuron circuit left and right members of each interneuron 
are grouped as one interneuron. 

\noindent 2) Synaptic and gap junction strengths between any two neurons are 
proportional to the anatomical number of synapses and gap junctions between them. 

\noindent 3) Pre-motor neurons do not generate sodium-type action potentials but 
their activities are graded, as {\it C. elegans} genome lacks molecules coding 
for voltage-activated sodium channels \cite{bargmann}. This assumption is 
also consistent with electrophysiological observations in {\it C. elegans} and 
related nematodes \cite{davis,goodman}.

\noindent 4) Each neuron releases a single neurotransmitter, or equivalently,
there exists a dominant neurotransmitter type for each neuron. Thus, with each
neuron in the command circuit we can associate a single dominant synaptic polarity.

\noindent 5) Worm's movement direction is determined by a relative imbalance in 
the activities of excitatory motor neurons of type A and B, which is in agreement 
with recent experimental observations \cite{kawano}.

\noindent 6) Behavioral output of the worm can be formally described in terms
of a three-state model. The three states correspond to forward motion, backward 
motion, and stopped period. Each state has its probability of occurrence, which 
for forward and backward states is given by an exponential function of a difference 
between activities of type A and B motor neurons (see below).

\vspace{0.45cm}

\noindent {\bf Derivation of the model equations.} 
Equations describing interneuron circuit responsible for forward-backward
motion transitions are given below. This is a nonlinear model based on
synaptic connectivity data from www.wormatlas.org (updated version of
White et al \cite{white} wiring diagram from \cite{chen}).

We start with a standard membrane equation describing the graded dynamics of neuron 
with an index $i$ \cite{ermentrout}:

\begin{eqnarray}
CS\frac{d V_{i}}{dt}= -g_{L}S(V_{i}-V_{r}) - \sum_{j} g_{s,ij}{\bf H_{0}}(V_{j})
(V_{i}-V_{s,j})   \nonumber  \\  
- \sum_{j} g_{e,ij}(V_{i}-V_{j})  
\end{eqnarray}\\
where $V_{i}$ is the voltage of neurons $i$, $C$ is the membrane capacitance per 
surface area, $S$ is the total surface area of neuron $i$, $V_{r}$ is the resting 
voltage, $g_{L}$ is the total membrane ionic conductance per surface area
that is composed mainly of a constant leak current (typical K$^{+}$ channel
conductance is much smaller for voltages close to $V_{r}$), $g_{e,ij}$ is the gap 
junction conductance between neurons $i$ and $j$. The symbol $g_{s,ij}$ denotes 
synaptic conductance coming from $j$ presynaptic neuron with synaptic reversal 
potential $V_{s,j}$. The function ${\bf H_{0}}(V_{j})$ is a nonlinear sigmoidal 
function characterizing synaptic transmission, and is given by 

\begin{eqnarray}
{\bf H_{0}}(V_{j})= \frac{1}{1 + \exp[-\gamma(V_{j}-\theta_{0})]},
\end{eqnarray}\\
where $\theta_{0}$ is the voltage threshold for synaptic activation, and 
$\gamma$ is a measure of steepness of the activation slope. Generally,
the synaptic input strongly depends on $\gamma$. We set its value at 0.15 
mV$^{-1}$ following an earlier analysis \cite{wicks} (Table 1). This particular 
value yields realistic synaptic currents that cause changes in voltage membrane
by at most several mV, which agrees with known neurophysiology in other animals
\cite{koch}.

Our goal is to write Eq. (1) in a more convenient form for the investigation
of synaptic polarities. We assume that the resting potential $V_{r}$ (when no
synaptic or gap junction input is present) for {\it C. elegans} interneurons 
is $-40$ mV, which agrees with earlier suggestions \cite{wicks}, and it is close
to a recent measurement ($\approx -50$ mV) in AIB neuron \cite{piggott}. We want to
re-define the voltage in Eq. (1) as a deviation from its resting value, i.e., we 
introduce $\Delta V_{i}\equiv V_{i}-V_{r}$. Let's denote by $V_{ex}$ the reversal 
potential for excitatory, and by $V_{in}$ the reversal potential for inhibitory 
synapses. The value of $V_{ex}$ is around 0 mV (the current in excitatory synapses 
is mediated by Na$^{+}$, K$^{+}$, and partly by Ca$^{++}$). The value of $V_{in}$ 
was reported between $-70$ mV \cite{purves} and $-90$ mV \cite{koch} (the current 
in inhibitory synapses is mediated by Cl$^{-}$). As an average for $V_{in}$ we take 
$-80$ mV. Consequently, we obtain for excitatory synapses 
$V_{i}-V_{ex}= \Delta V_{i} + V_{r}-V_{ex}= \Delta V_{i} - 40$,
and for inhibitory
$V_{i}-V_{in}= \Delta V_{i} + V_{r}-V_{in}= \Delta V_{i} + 40$. The resulting 
average numerical factors in both expressions have identical absolute values. 
Thus we can use an approximation: 
$V_{i}-V_{s,j}\approx \Delta V_{i} - \epsilon_{j}A$, where $A= 40$ mV, and 
$\epsilon_{j}$ characterizes the synaptic polarity of the presynaptic neuron
$j$. The value of $\epsilon_{j}$ is either 1 for excitatory synapses or $-1$ 
for inhibitory. Taking the above into account and dividing both sides of Eq. (1) 
by $g_{L}S$, we can rewrite this equation as

\begin{eqnarray}
\frac{C}{g_{L}}\frac{d(\Delta V_{i})}{dt}= -\left[1 + 
\sum_{j}\frac{g_{s,ij}{\bf H_{0}}(V_{j})}{g_{L}S}\right]\Delta V_{i}   
+ \sum_{j} \epsilon_{j}A \frac{g_{s,ij}}{g_{L}S}{\bf H_{0}}(V_{j}) 
 \nonumber  \\  
- \sum_{j} \frac{g_{e,ij}}{g_{L}S}(\Delta V_{i}- \Delta V_{j}). 
\end{eqnarray}\\
We can determine the strengths of synaptic and gap junction connections 
between any $i$ and $j$ interneurons by their anatomical numbers 
$N_{s,ij}, N_{e,ij}$, and maximal elementary conductances $q_{s}, q_{e}$, 
i.e., $g_{s,ij}= N_{s,ij}q_{s}$, and $g_{e,ij}= N_{e,ij}q_{e}$. The data for 
$N_{s,ij}$ and $N_{e,ij}$ are available from the data set in \cite{chen} 
(see Table 7). A typical range of conductances for chemical and electrical 
synapses is known from neurophysiology of other animals \cite{koch}. 
The leak conductance $g_{L}$ is taken as
$g_{L}^{-1}= 150$  $k\Omega\cdot cm^{2}$ \cite{wicks}, which comes form the 
neurophysiological measurements in a related larger nematode {\it Ascaris}
\cite{davis}. The surface area $S$ of all interneurons is very similar
and around $15\cdot 10^{-6}$ cm$^{2}$ \cite{white,wicks}, so we obtain
$g_{L}S= 0.1$ nS, of which the inverse (i.e. $10^{10}$ $\Omega$) is comparable
to an experimental measurement of the total input resistance $\sim 0.5\cdot 
10^{10}$ $\Omega$ \cite{goodman}. 
Consequently, we can estimate the ratios present in Eq. (3) as: 
$A g_{s,ij}/(g_{L}S)= 400 N_{s,ij}q_{s}$, and  
$g_{e,ij}/(g_{L}S)= 10 N_{e,ij}q_{e}$, where $q_{s}, q_{e}$ are expressed in nS. 
We checked that the term $\sum_{j} g_{s,ij}{\bf H}(V_{j})/(g_{L}S)$, 
which is associated with $\Delta V_{i}$ is generally much smaller than 1, 
since ${\bf H}(V_{j}) \ll 1$ for voltages not far away from $V_{r}$. 
Consequently, this term is neglected, which simplifies significantly the 
resulting equations for interneurons (see below and the Supplementary Information). 
Thus, we can write Eq. (3) in an approximate form as:

\begin{eqnarray}
\tau \frac{d(\Delta V_{i})}{dt}\approx - \Delta V_{i}   
+ \sum_{j} \epsilon_{j}w_{ij}{\bf H}(\Delta V_{j}) 
\nonumber  \\  
- \sum_{j} g_{ij}(\Delta V_{i}- \Delta V_{j}),
\end{eqnarray}\\
where $\tau= C/g_{L}$ is the membrane time constant, $w_{ij}$ is the synaptic
coupling $w_{ij}= 400q_{s} N_{s,ij}$, and $g_{ij}$ is the gap junction coupling
$g_{ij}= 10q_{e} N_{e,ij}$. The function
${\bf H}(\Delta V_{j})$ in Eq. (4) differs from the original function 
${\bf H_{0}}(V_{j})$ only by a substitution $\theta_{0} \rightarrow \theta$,
where $\theta= \theta_{0} - V_{r}$, i.e.

\begin{eqnarray}
{\bf H}(\Delta V_{j})= \frac{1}{1 + \exp[-\gamma(\Delta V_{j}-\theta)]}.
\end{eqnarray}

It is important to keep in mind that synaptic polarities in Eq. (4) are
determined simply by the signs of $\epsilon_{j}$ coefficients.

\vspace{0.45cm}

\noindent {\bf Activity equations for interneurons.} Equations describing
activities of the interneurons in the locomotory circuit are presented in 
the Supplementary Information. They are similar in form to Eq. (4) with an 
additional inclusion of the heterogeneous excitatory sensory input $X_{i}$, i.e.

\begin{eqnarray}
\tau \frac{d(\Delta V_{i})}{dt}= - \Delta V_{i}   
+ \sum_{j} \epsilon_{j}w_{ij}{\bf H}(\Delta V_{j}) 
\nonumber  \\  
+ \sum_{j} \epsilon_{i}^{2}\epsilon_{j}^{2} 
g_{ij}(\Delta V_{j}- \Delta V_{i}) + X_{i},
\end{eqnarray}\\
where the subscripts $i, j$ are labels for our circuit neurons. The parameter 
$\epsilon_{i}$ denotes synaptic polarity of the neuron $i$, and it assumes value 
1 (if the neuron is excitatory), value $-1$ (if inhibitory), or 0 (if the neuron 
is absent because of the ablation). Ablations in the circuit remove also electric
connections if a neuron on either side of this coupling is killed. To include this 
effect we rescale the gap junction couplings by $\epsilon_{i}^{2}\epsilon_{j}^2$ 
factors. The square in the epsilon assures that we do not get unphysical negative 
values for gap junction conductance. We assume that the excitatory sensory input 
$X_{i}$ coming from the upstream neurons to the interneuron $i$ is constant in time 
and represented by 

\begin{equation}
X_{i}= x_{0} + \sigma z_{i}, 
\end{equation}
where $x_{0}= 2.0$ mV, $\sigma$ is a variable parameter characterizing the strength
of a strong input that is constant in time, and $z_{i}$ is the binary variable either 
0 (weak input) or 1 (strong input). Thus neurons can receive only two types of the 
input: either weak ($x_{0}$) or strong ($x_{0}+\sigma$). The ``input'' parameter 
$z_{i}$, similar to $\epsilon_{i}$, is unknown. We want to find their optimal values 
for each interneuron.

The above pre-motor interneurons make synaptic and gap junction connections with
downstream excitatory motor neurons. Two separate groups of these motor neurons
generating forward and backward motion, called B and A respectively, directly 
connect locomotory muscles. The activities of excitatory motor neurons ($E_{f}$
and $E_{b}$ for forward and backward motion respectively) have a similar form 
to that of the interneurons and are given in the Supplementary Information.
Generally, activity equations for interneurons and motor neurons are of similar 
kind to those used before in \cite{karb2008} for analyzing forward locomotion.

We solve Eqs. (6-7) using a second order Runge-Kutta method. We are interested only 
in the steady-state activities of the interneurons and motor neurons. One can think 
about these activities as temporal averages over sufficiently long periods of time
driven by a constant bias input. This simplifying step significantly enhances 
the feasibility of the analysis. The steady-state values of 
motor neuron activities, $E_{f}$ and $E_{b}$, are inserted in Eq. (8); see below.
All the results presented were obtained by choosing initial conditions corresponding
to resting voltages, i.e. $\Delta V_{i}= 0$ for each neuron. We also tried random
initial conditions in which neurons start with voltages uniformly distributed
in the range $-10$ to 10 mV. In both cases the steady-state values are the same,
which means that steady-state activities are indepedent of initial conditions, at
least in that range.

\vspace{0.45cm}

\noindent {\bf Theoretical ablations.} Ablations or removals of neurons in the 
model are performed by setting $\epsilon_{neuron}= 0$. For example, if we remove 
neuron AVB, then we put $\epsilon_{AVB}= 0$ in all equations for neural activities.

\vspace{0.45cm}

\noindent {\bf Values of the connectivity matrix}. The strength $w_{ij}$ of 
synaptic input to neuron $i$ coming from neuron $j$ is given according 
to Eq. (4) by the expression $w_{ij}= 400q_{s} N_{s,ij}$, where $N_{s,ij}$ is 
the number of synaptic contacts of neuron $i$ with presynaptic neuron $j$. 
The number $N_{s,ij}$ is determined as an arithmetic mean for the right- and 
left-hand side interneurons. As an example, the right AVB neuron receives an input
from both right and left PVC neurons, of which we take an arithmetic mean.
Similarly, the left AVB neuron receives an input from both right and left PVC, 
of which we again take an arithmetic mean. Next, we take an arithmetic
mean of these two arithmetic means, and obtain a single value representing
average number of synaptic contacts between presynaptic PVC and postsynaptic AVB
($N_{s,AVB,PVC}$). The strength of gap junction $g_{ij}$ between neurons $i$ and $j$ 
is given by an analogous formula $g_{ij}= 10q_{e} N_{e,ij}$, where $N_{e,ij}$ is the 
number of gap junction contacts between $i$ and $j$ (arithmetic mean of right and 
left interneurons). Empirical data for $N_{s,ij}$ and $N_{e,ij}$ were taken from the 
database in \cite{chen} and are presented in Table 7. Parameters $q_{s}$ and $q_{e}$ 
were taken in the range: $q_{s}= 0.03-0.6$ nS, and $q_{e}= 0.03-0.5$ nS \cite{koch}.

\vspace{0.45cm}

\noindent {\bf ASH neuron.} From all upstream neurons we selected explicitly
ASH neuron because of its polymodal sensory role. Specifically, it has been 
implicated in avoidance responses \cite{kaplan}, which are associated with 
the modulation of locomotion. We do not write an explicit equation for ASH dynamics, 
because it receives a massive input form many other head neurons, of which 
we have no knowledge. Instead, we make computations for 4 different levels of 
ASH activity that we set by hand. We choose $ASH= \kappa\theta$, 
where $\kappa= 0.2$, 0.4, 0.6, or 0.75. The value of the normalized threshold 
$\theta$ is set to 45 mV.

\vspace{0.45cm}

\noindent {\bf Relationship to the behavioral data.} From the experimental 
part we have average times the worms spent in forward and backward locomotion, 
which we denote as $T_{f}$ and $T_{b}$, respectively. These average times should 
be somehow related to the average activities of the two type of motor neurons, 
$E_{f}$ and $E_{b}$. The precise relationship between them is unknown due to 
the lack of direct physiological data. However, one can expect that domination
of $E_{f}$ over $E_{b}$ should favor forward motion and its duration, and vice
versa.

We can make some progress by using an analogy with statistical physics 
\cite{gardiner}, and treating worm's locomotory behavior as a three state system 
influenced by both deterministic and stochastic factors. These three states 
correspond to forward movement, backward movement, and stopped time (no motion). 
There could be transitions between the states driven by
sensory input from the environment (either deterministic or stochastic).
However, we do not model such transitions. We are interested only in long-term
``average'' or steady states activities of the system.
Our model is motivated in large part by experimental results of Kawano et al 
\cite{kawano}. In that study it was shown that {\it C. elegans} motion direction
is determined by a relative activity of A and B motor neurons. In particular,
it was suggested \cite{kawano} that when forward motor neurons are much more 
active than backward (i.e. $E_{f} \gg E_{b}$), then there should be a high 
probability of finding the worm in the forward motion. Conversely, if the activity 
of backward motor neurons dominates over the activities of their forward counterparts
(i.e. $E_{b} \gg E_{f}$), then there is a high chance that the worm moves backward. 
Thus, it appears that the sign of $E_{f}-E_{b}$ plays a key role in the choice of 
worm's motion direction. Moreover, one can expect that when the activities
of both types of motor neurons are comparable (equal or almost equal), then
{\it C. elegans} likely does not move.

We relate the behavioral observables, i.e. the fraction of time spent in
forward motion, with the motor neuron activities using a transfer function
known as a sigmoidal logistic function. Specifically, we propose

\begin{equation}
\frac{T_{f}}{(T_{f}+T_{b})}= \frac{1}{1 + \exp[(E_{b}-E_{f})/\eta]},
\end{equation}\\
where $\eta$ is the noise in the system. This parameter also determines
the shape of the transfer function. This type of function is a standard tool 
often used in many computational studies in neuroscience, and this is the 
primary transfer function used in this study. Note that for cases in which
activities of A motor neurons dominate, i.e. $(E_{b}-E_{f})/\eta \gg 1$, 
the time spent in forward motion is relatively short, i.e. $T_{f}/T_{b} \ll 1$.
The derivation of Eq. (8) is presented in the Supplementary Information.

\vspace{0.45cm}

\noindent {\bf The goal.} We want to determine which combination of neuron 
polarities: $\epsilon_{ASH}$, $\epsilon_{AVB}$, $\epsilon_{PVC}$, 
$\epsilon_{DVA}$, $\epsilon_{AVA}$, $\epsilon_{AVD}$, $\epsilon_{AVE}$,
together with their corresponding upstream inputs $z_{i}$, yields the 
best fit to the experimental values of $T_{f}/(T_{f}+T_{b})$.

\vspace{0.45cm}

\noindent {\bf Comparison of the theory with the data.} In our model there 
are 8192 distinct combinations of synaptic polarities $\epsilon_{i}$ and 
the upstream inputs $z_{i}$, i.e., different configurations in which the 
circuit can be found. For each circuit configuration, we perform 17 interneuron 
ablations in our computer model, and compute theoretical values of 
$T_{f}/(T_{f}+T_{b})$ for each ablation. Next, we compute an Euclidean distance 
ED of these values to the experimental values given in Table 2. The Euclidean
distance serves as a system performance (the lower ED the better), and it is
computed according to the expression:

\begin{eqnarray}
\mbox{ED}= \left[ \sum_{a=1}^{18} (R_{th}-R_{ex})_{a}^{2}\right]^{1/2},
\end{eqnarray}\\
where 
$R= T_{f}/(T_{f}+T_{b})\equiv (T_{f}/T_{b})\left[1+ T_{f}/T_{b}\right]^{-1}$, 
and the subscripts $th$ and $ex$
refer to theoretical and experimental values of $R$. The subscript $a$ 
refers to the ablation number, in the same order as in Table 2.
In particular, $a= 1$ corresponds to the mock ablation, i.e. wild type (WT).
All the results presented in Tables 3-6 were generated by using Eq. (8) for the
parameter $R_{th}$.

\vspace{1.3cm}

\noindent{\bf Acknowledgments}

We thank Cori Bargmann for comments on the manuscript.
The work was supported by the EU grant POIG.02.01.00-14-122/09 ``Physics as 
the basis for new technologies - development of modern research infrastructure
at the Faculty of Physics of the University of Warsaw'' (FR), by the Marie Curie 
Actions EU grant FP7-PEOPLE-2007-IRG-210538 (JK), and by the grant from the 
Polish Ministry of Science and Education NN 518 409238 (JK).

\newpage

\vspace{1.5cm}



\newpage

{\bf \large Figure Captions}

Fig. 1\\
Schematic diagram of the interneuron locomotory circuit.
(A) Intact circuit. ASH neuron is an upstream neuron that provides 
synaptic input to the locomotory interneurons. The output coming from 
the 6 neurons (5 interneurons and DVA) feeds the activities of motor neurons, 
represented by $E_{f}$ (controlling forward motion) and by $E_{b}$ (controlling 
backward motion). Synaptic connections are shown as solid arrows (blue),
and gap junctions are represented by dashed lines (red). The magnitude of 
an arrow and the width of a dashed line are indicators of the strength 
of synaptic and gap junction connections, respectively.  
(B) An example of an ablated circuit, in which ASH and AVB neurons are 
removed. Note that this leads to the removal of all connections (synaptic 
and electric) coming out from these neurons. Such ablations not only change 
the circuit architecture but also modify its activity output.

\vspace{0.3cm}

Fig. 2\\
Comparison of the theory with the data for relative times spent in forward 
locomotion across different ablations. The theoretical values are for the
winning polarity combination $\#$ 1, corresponding to all inhibitory neurons.
Correlation between theoretical points (red triangles) and experimental
(blue circles) is relatively high ($R=0.743$) and statistically significant
($p=0.0004$). The error bars for the experimental points were computed
from SEM values of $T_{f}$ and $T_{b}$ given in Table 2.
The optimal values of the free parameters are given in Table 3.

\vspace{2.3cm}

Fig. 3\\
Dependence of the Euclidean Distance (ED) on the patterns of synaptic
polarities and input strength. Neurons receiving a strong input are marked
in orange. Inhibitory neurons are represented in red, while excitatory
in blue. The smallest (optimal) value of ED is pinpointed by a pink arrow.
Note that configurations with small ED are generally associated with
mostly inhibitory connections and a moderate input strength (left part
of the map), whereas large ED values occur for mostly excitatory configurations
(right part of the map). The optimal parameters are the same as in Fig. 2.

\vspace{0.3cm}

Fig. 4\\
Distribution of synaptic polarities for each interneuron. The first 
20 polarity combinations with the smallest Euclidean distance (ED) are
shown, and they are associated with the optimal parameters given in Table 3. 
Note that the interneurons ASH, AVA, AVE, and PVC are inhibitory with a high
probability. There are some nonzero likelihoods that AVB, AVD, and DVA neurons 
are excitatory (especially AVB and AVD), although the smallest ED values are 
associated with negative polarities. The optimal parameters are the same as in
Fig. 2.

\vspace{0.3cm}

Fig. 5\\
Dependence of ED on synaptic and electric conductances. ED is optimal
(minimal) for some range of values of $q_{s}$ and $q_{e}$. Note that
the changes in synaptic conductance (horizontal direction) are more critical 
for ED than the changes in $q_{e}$ (vertical direction). The optimal values
of other parameters are: $\sigma= 8.0$ mV, $\kappa= 0.6$, and $\eta= 1.05$ mV.

\vspace{0.3cm}

Fig. 6\\
Dependence of ED on the system noise amplitude $\eta$.
ED has a minimum for some optimal $\eta$, and this is essentially independnet
of the synaptic polarity configuration. Shown are synaptic configurations 
number 1 (solid line), 17 (dashed line), and 11 (dotted line). The optimal
values of other parameters are: $\sigma= 8.0$ mV, $\kappa= 0.6$, and 
$q_{s}= q_{e}= 0.1$ nS.

\newpage

\begin{table}
\begin{center}
\caption{The main parameters used in the model.}
\begin{tabular}{|l l l|}
\hline

Symbol  &   Value or Range   &   Description    \\

\hline

$q_{s}$   & 0.03-0.6 nS (optimized)  &   Maximal conductance of a single synapse   \\
$q_{e}$   & 0.03-0.5 nS (optimized)  &   Conductance of a single gap junction   \\
$\theta$  & 45 mV (fixed)            &   Renormalized threshold for synaptic activation  \\
$\gamma$  & 0.15 mV$^{-1}$  (fixed)  &   Steepness of synaptic activation   \\
$\sigma$  & 4-12 mV (variable)       &   Amplitude of strong input    \\
$z_{i}$   &    0 or 1                &   Binary variable indicating the presence of strong input  \\
$\kappa$  & 0.2-0.75  (variable)     &   Excitation level of ASH neuron   \\
$\eta$    & 0.1-2.0 mV (optimized)   &   Noise in the system    \\

\hline


\end{tabular}
\end{center}

\end{table}

\newpage

\begin{table}
\begin{center}
\caption{Experimental data of the impact of neuron ablation on {\it C. elegans}
locomotion. Shown are population average times and standard errors of the mean
(SEM) for worms in forward and backward motion, and in stopped phase. The last 
column gives population averages of reversal frequencies with their SEM.}
\begin{tabular}{|l l l l l|}
\hline

Ablation type &  Forward time   &  Backward time  &  Stopped    & Reversals   \\
              &   $T_{f}$ (sec) &   $T_{b}$ (sec) &  time (sec) & (min$^{-1}$) \\

\hline

mock ablated (WT, N=43) &  8.98 $\pm$ 0.57  &  2.80 $\pm$ 0.27  &  0.26 $\pm$ 0.01  &   5.29 $\pm$ 0.27   \\
ASH (N=14)              &  12.6 $\pm$ 1.67  &  0.93 $\pm$ 0.17  &  0.27 $\pm$ 0.01  &   3.79 $\pm$ 0.80   \\
AVA (N=11)              &  0.71 $\pm$ 0.09  &  0.53 $\pm$ 0.04  &  0.60 $\pm$ 0.05  &   10.3 $\pm$ 0.56   \\
AVB (N=8)               &  2.26 $\pm$ 0.40  &  2.14 $\pm$ 0.23  &  0.38 $\pm$ 0.02  &   6.10 $\pm$ 0.64   \\
AVD (N=4)               &  4.23 $\pm$ 1.80  &  3.12 $\pm$ 0.36  &  0.31 $\pm$ 0.04  &   3.50 $\pm$ 0.31   \\
DVA (N=22)              &  1.51 $\pm$ 0.18  &  1.23 $\pm$ 0.08  &  0.44 $\pm$ 0.02  &   10.0 $\pm$ 0.57   \\
PVC (N=12)              &  12.0 $\pm$ 1.81  &  1.89 $\pm$ 0.39  &  0.29 $\pm$ 0.02  &   5.46 $\pm$ 0.74   \\
ASH+AVA (N=7)           &  1.91 $\pm$ 0.42  &  0.85 $\pm$ 0.20  &  0.52 $\pm$ 0.06  &   5.18 $\pm$ 0.90   \\
ASH+AVB (N=12)          &  2.05 $\pm$ 0.47  &  2.04 $\pm$ 0.43  &  0.42 $\pm$ 0.06  &   6.92 $\pm$ 1.08   \\
AVA+AVB (N=9)           &  0.56 $\pm$ 0.14  &  0.46 $\pm$ 0.06  &  0.89 $\pm$ 0.17  &   10.1 $\pm$ 1.36   \\
AVA+PVC (N=11)          &  4.09 $\pm$ 0.91  &  0.67 $\pm$ 0.14  &  0.37 $\pm$ 0.04  &   9.78 $\pm$ 0.71    \\
AVB+PVC (N=5)           &  0.91 $\pm$ 0.24  &  1.19 $\pm$ 0.19  &  0.44 $\pm$ 0.08  &   15.0 $\pm$ 3.52   \\
DVA+PVC (N=19)          &  2.18 $\pm$ 0.21  &  1.35 $\pm$ 0.08  &  0.40 $\pm$ 0.02  &   11.6 $\pm$ 0.57   \\
ASH+AVA+AVB (N=8)       &  0.75 $\pm$ 0.24  &  0.52 $\pm$ 0.11  &  1.16 $\pm$ 0.26  &   6.47 $\pm$ 0.83    \\
AVA+AVB+PVC (N=8)       &  0.93 $\pm$ 0.33  &  0.47 $\pm$ 0.12  &  0.87 $\pm$ 0.21  &   6.10 $\pm$ 0.87   \\
AVB+AVD+PVC (N=5)       &  1.33 $\pm$ 0.31  &  0.94 $\pm$ 0.13  &  0.49 $\pm$ 0.07  &   11.2 $\pm$ 2.00   \\
AVB+DVA+PVC (N=10)      &  1.90 $\pm$ 0.28  &  1.03 $\pm$ 0.14  &  0.40 $\pm$ 0.21  &   12.2 $\pm$ 1.53    \\
AVA+AVB+AVE+PVC (N=10)  &  0.60 $\pm$ 0.21  &  0.39 $\pm$ 0.14  &  1.00 $\pm$ 0.12  &   8.66 $\pm$ 1.54    \\

\hline

\hline
\end{tabular}
\end{center}
\end{table}

\newpage

\begin{table}
\begin{center}
\caption{The winning combinations of interneuron polarities ($\epsilon_{i}$)
for the upstream input: $\sigma= 8.0$ mV and $\kappa= 0.6$. }
\begin{tabular}{|l | l l l l l l l l | l|}
\hline

Neuron  &  \multicolumn{8}{|c|}{Rank}  &   inhibitory    \\
        &  1  &  2  &  3  &  4  &  5  &  6  &  7  &  8  &  likelihood    \\

\hline

ASH     &   -1   &   -1   &   -1  &   -1   &   -1   &  -1  &  -1  &  -1  &  1   \\     
AVA     &   -1   &   -1   &   -1  &   -1   &   -1   &  -1  &   1  &   1  &  3/4  \\     
AVB     &   -1   &    1   &   -1  &    1   &    1   &  -1  &  -1  &  -1  &  5/8   \\     
AVD     &   -1   &   -1   &    1  &    1   &   -1   &  -1  &  -1  &  -1  &  3/4   \\     
AVE     &   -1   &   -1   &   -1  &   -1   &   -1   &  -1  &  -1  &  -1  &  1     \\     
DVA     &   -1   &   -1   &    1  &    1   &    1   &   1  &   1  &  -1  &  3/8   \\     
PVC     &   -1   &   -1   &   -1  &   -1   &   -1   &  -1  &  -1  &  -1  &  1    \\     

\hline

Combination $\#$ &   1  &  17  &  11  &  27  &  19   &   3  & 35  &  33    \\     
ED     & 0.3625 & 0.3651 & 0.374 & 0.377 & 0.380 & 0.383 & 0.396 &  0.409  \\
Corr   & 0.7433 & 0.7417 & 0.722 & 0.717 & 0.740 & 0.746 & 0.690 &  0.731  \\


\end{tabular}
\end{center}
The optimal values of the parameters are: $\eta= 1.05$ mV, $q_{s}= 0.1$ nS, 
$q_{e}= 0.1$ nS. All the combinations receive the same ``winning'' input: 
$z_{AVB}= 1$, $z_{PVC}= 1$, and $z_{i}= 0$ for other interneurons. 
The last column provides estimates of probabilities that a given neuron is 
inhibitory. In all Tables 3-6, ED was computed using Eqs. (8) and (9).
\end{table}

\newpage

\begin{table}
\begin{center}
\caption{The winning combinations of interneuron polarities ($\epsilon_{i}$)
for the upstream input: $\sigma= 6.0$ mV and $\kappa= 0.6$. }
\begin{tabular}{|l | l l l l l l l l | l |}
\hline

Neuron  &  \multicolumn{8}{|c|}{Rank}  &   inhibitory    \\
        &  1  &  2  &  3  &  4  &  5  &  6  &  7  &  8  &  likelihood    \\

\hline

ASH     &   -1   &   -1   &   -1  &   -1   &   -1   &    -1   &  -1  &  -1  &  1  \\     
AVA     &   -1   &   -1   &   -1  &   -1   &   -1   &    -1   &  -1  &  -1  &  1   \\     
AVB     &   -1   &    1   &   -1  &    1   &   -1   &     1   &   1  &   1  &  3/8  \\     
AVD     &   -1   &   -1   &    1  &    1   &   -1   &    -1   &   1  &   1  &  1/2   \\     
AVE     &   -1   &   -1   &   -1  &   -1   &   -1   &    -1   &  -1  &  -1  &  1    \\     
DVA     &   -1   &   -1   &    1  &    1   &    1   &     1   &  -1  &   1  &  1/2   \\     
PVC     &   -1   &   -1   &   -1  &   -1   &   -1   &    -1   &   1  &  -1  &  7/8   \\     

\hline

Combination $\#$ &   1  &  17  &  11  &  27  &  3   &   19  &  26    &  10   \\     
ED     & 0.368 & 0.377 & 0.394 & 0.404 & 0.422 &  0.424  &  0.431 & 0.433   \\     
Corr   & 0.734 & 0.722 & 0.687  & 0.665 & 0.692 & 0.669  & 0.627  &  0.637    \\


\end{tabular}
\end{center}
The optimal values of the parameters are: $\eta= 0.85$ mV, $q_{s}= 0.1$ nS, 
$q_{e}= 0.1$ nS. All the combinations receive the same ``winning'' input:
$z_{AVB}= 1$, $z_{PVC}= 1$, and $z_{i}= 0$ for other interneurons. 
The last column provides estimates of probabilities that a given neuron is inhibitory. 

\end{table}

\newpage

\begin{table}
\begin{center}
\caption{The winning combinations of interneuron polarities ($\epsilon_{i}$)
for the upstream input: $\sigma= 4.0$ mV and $\kappa= 0.6$. }
\begin{tabular}{|l | l l l l l l l l | l |}
\hline

Neuron  &  \multicolumn{8}{|c|}{Rank}  &   inhibitory    \\
        &  1  &  2  &  3  &  4  &  5  &  6  &  7  &  8  &  likelihood    \\

\hline

ASH     &   -1   &   -1   &   -1  &   -1   &   -1   &  -1   &  -1   &  -1 &  1     \\     
AVA     &   -1   &   -1   &   -1  &   -1   &   -1   &  -1   &  -1   &  -1 &  1   \\     
AVB     &   -1   &    1   &   -1  &   -1   &   -1   &   1   &   1   &   1 &  1/2  \\     
AVD     &   -1   &   -1   &    1  &    1   &    1   &   1   &   1   &   1 &  1/4  \\     
AVE     &   -1   &   -1   &   -1  &   -1   &   -1   &  -1   &  -1   &  -1 &  1   \\     
DVA     &   -1   &   -1   &    1  &   -1   &   -1   &   1   &  -1   &  -1 &  3/4  \\     
PVC     &   -1   &   -1   &   -1  &   -1   &    1   &  -1   &   1   &  -1 &  3/4  \\     

\hline

Combination $\#$ &   1  &  17  &  11  &  9  &  10  &  27  &  26  &  25    \\     
ED     & 0.414 & 0.431 & 0.453 & 0.461 & 0.465 & 0.471 &  0.476 &  0.487 \\
Corr   & 0.644 & 0.606 & 0.560 & 0.613 & 0.530 & 0.512 &  0.485 &  0.578 \\


\end{tabular}
\end{center}
The optimal values of the parameters are: $\eta= 0.7$ mV, $q_{s}= 0.1$ nS, 
$q_{e}= 0.1$ nS. All the combinations receive the same ``winning'' input:
$z_{AVB}= 1$ and $z_{i}= 0$ for other interneurons. The last column 
provides estimates of probabilities that a given neuron is inhibitory. 

\end{table}

\newpage

\begin{table}
\begin{center}
\caption{The winning combinations of interneuron polarities ($\epsilon_{i}$)
for the upstream input: $\sigma= 12.0$ mV and $\kappa= 0.6$. }
\begin{tabular}{|l | l l l l l l l l | l |}
\hline

Neuron  &  \multicolumn{8}{|c|}{Rank}  &   inhibitory    \\
        &  1  &  2  &  3  &  4  &  5  &  6  &  7  &  8  &  likelihood    \\

\hline

ASH     &   -1   &   -1   &   -1  &   -1   &   -1   &  -1  &  -1 &  -1  &  1   \\     
AVA     &   -1   &   -1   &    1  &    1   &   -1   &  -1  &  -1 &  -1  &  3/4  \\     
AVB     &    1   &   -1   &    1  &   -1   &    1   &  -1  &  -1 &   1  &  1/2  \\     
AVD     &    1   &    1   &    1  &    1   &   -1   &  -1  &  -1 &  -1  &  1/2  \\     
AVE     &   -1   &   -1   &   -1  &   -1   &    1   &   1  &  -1 &  -1  &  3/4  \\     
DVA     &   -1   &   -1   &   -1  &   -1   &   -1   &  -1  &  -1 &  -1  &  1    \\     
PVC     &    1   &    1   &    1  &    1   &    1   &   1  &  -1 &  -1  &  1/4   \\     

\hline

Combination $\#$ &  26  &  10  &  58  &  42  &  22  &  6  &  1  &  17     \\     
ED     & 0.383 & 0.386 & 0.390 & 0.392 & 0.393 & 0.394  &  0.397  & 0.398     \\     
Corr   & 0.715 & 0.708 & 0.694 & 0.688 & 0.691 & 0.686  & 0.687   & 0.685   \\


\end{tabular}
\end{center}
The optimal values of the parameters are: $\eta= 0.85$ mV, $q_{s}= 0.1$ nS, 
$q_{e}= 0.3$ nS. All the combinations receive the same ``winning'' input: 
$z_{AVB}= 1$, $z_{PVC}= 1$, and $z_{i}= 0$ for other interneurons. 
The last column provides estimates of probabilities that a given neuron is inhibitory. 

\end{table}

\newpage

\begin{table}
\begin{center}
\caption{Connectivity matrix for the command interneuron circuit. 
Shown are average anatomical number of synapses $(N_{s,ij})$ and gap junctions 
$(N_{e,ij})$ (in the brackets below synaptic contacts) between postsynaptic neuron 
$i$ and presynaptic neuron $j$. Symbols F and B denote forward and backward motor 
neurons, respectively. }
\begin{tabular}{|l | l l l l l l l l l|}

\hline
postsynaptic   &  \multicolumn{9}{c|}{presynaptic neurons}         \\  
neuron             &   &   &   &   &   &   &   &   &       \\  

               & ASH  & AVA  & AVB  & AVD  & AVE  & DVA  & PVC  &  F &  B    \\

\hline

AVA    &  1.75 &  $-$ & 6.75 & 15.75 & 10.5 &  2.0  &  5.0  & $-$   &  0.25     \\     
       &  $-$  &  $-$ & $-$  & $-$   & $-$  &  $-$  & (2.5) & (3.5) &  (25.5)   \\
       &       &      &      &       &      &       &       &       &          \\
AVB    &  2.25 &  0.5 & $-$ & 0.25   & $-$  &  0.5  &  7.75 & $-$   &  $-$      \\
       &  $-$  &  $-$ & $-$  & $-$   & $-$  & (1.0) &  $-$  & (13.75) & (0.5)    \\
       &       &      &      &       &      &       &       &       &            \\
AVD    &  3.0  &  1.0 &  0.75 & $-$  & 0.25 &  $-$  &  3.25 &  $-$  &  0.25      \\
       &  $-$  &  $-$ &  $-$  & $-$  & $-$  &  $-$  &  $-$  &  $-$  &  $-$      \\
       &       &      &       &      &      &       &       &       &            \\     
AVE    &  0.75 &  1.0 &  0.75 & $-$  & $-$  &  7.0  &  1.25 &  $-$  &  $-$       \\  
       &  $-$  &  $-$ &  $-$  & $-$  & $-$  &  $-$  &  $-$  &  $-$  &  $-$       \\
       &       &      &       &      &      &       &       &       &            \\
DVA    &  $-$  &  $-$ &  $-$  & $-$  & $-$  &  $-$  &  2.0  &  0.5  &  $-$       \\
       &  $-$  &  $-$ & (1.0) & $-$  & $-$  &  $-$  & (0.5) & (0.5) &  $-$       \\
       &       &      &       &      &      &       &       &       &            \\     
PVC    &  $-$  &  7.0 &  $-$  & 0.25 & 0.25 &  2.0  &  $-$  &  0.25 &  1.25      \\
       &  $-$  & (2.5) & $-$  & $-$  & $-$  & (0.5) & $-$  & (0.75) & (0.75)   \\ 
       &       &       &      &      &       &       &      &       &           \\    
F      &  $-$  &  2.5  & 0.25 &  0.25 & 0.25 &  6.5  & $-$  &  $-$  &  $-$      \\
       &  $-$  & (3.5) & (13.75)& $-$ & $-$  & (0.5) & (0.75) & $-$ &  $-$      \\
       &       &       &       &      &      &       &      &        &          \\
B      &  $-$  &  41.75 & 1.5  &  7.0 & 8.25 &  1.0  &  1.0  &  $-$  & $-$      \\
       &  $-$  & (25.5) & (0.5) & $-$ & $-$  &  $-$  & (0.75) & $-$  & $-$      \\

\hline

\hline
\end{tabular}
\end{center}

\end{table}

\end{document}